**Chapter 7**

**Vibrational dynamics of non-crystalline solids**

Giacomo Baldi, Aldo Fontana and Giulio Monaco

*Department of Physics, University of Trento,*
*Via Sommarive 14, 38123 Povo (Tn), Italy*
*giacomo.baldi@unitn.it*

The boson peak (BP) is an excess of vibrational states over the Debye law appearing at terahertz frequencies. It is found in all glasses and marks the crossover between the long-wavelength behavior, where the solid can be considered as an isotropic continuum, and the region where the wavelength of the sound wave starts to experience the microscopic details of the structure. This chapter is devoted to a review of the main experimental observations regarding the vibrational dynamics of amorphous solids, as detected by neutron and X-ray scattering techniques. A first part of the chapter is devoted to the measurements of the BP and its evolution as a function of external parameters, such as temperature, pressure, or density. The second part of the chapter reviews the wavevector evolution of the dynamic structure factor, which provides evidence of a pseudo-acoustic propagating mode up to frequencies comparable to those of the BP. This longitudinal mode has a sound attenuation which follows the Rayleigh scattering law and a negative dispersion that can partially explain the deviation from the Debye law. At higher frequencies, the inelastic spectrum presents a complex pattern of vibrations, with evidences of two peaks in various systems. To conclude, we will highlight the information that can be gained on the nature of the glass vibrational modes from a comparison with the dynamics of the corresponding polycrystal.

## 1. Introduction

At the macroscopic scale, a glass is well described as an isotropic continuum. Classical elasticity theory[1] correctly predicts the propagation of elastic waves in the long-wavelength limit, albeit the theory ignores the presence of internal stresses, natively stored in the material during its production. However, the density of vibrational states (DOS), measured by means of Raman or inelastic neutron scattering, deviates from the Debye prediction, whose validity would be expected for the sound waves (see chapter 1). At frequencies of a few terahertz the ratio between the density of states and the Debye law shows a peak, usually termed "boson peak" for reasons that will be clear in the next section. The nature of the vibrational modes associated with this peak is matter of active research. The typical frequency of the BP corresponds to wavelengths of tens of interatomic spacings, meaning that the microscopic structure of the solid cannot be neglected in the description of the vibrational states. The determination of the vibrational modes of a disordered solid is significantly more complex than for a single crystal because the absence of translational periodicity precludes the use of concepts such as the reciprocal space and the Bloch theorem. For this reason, many models have been proposed but a microscopic description of the terahertz vibrations of glasses is still missing.

Without attempting to present a comprehensive account of all the various theoretical approaches, we will here mention a few of them. The soft potential model[2] is based on an extension of the concept of defects with two-level states[3]. It assumes the presence of soft potentials and of quasi-localized excitations in addition to the sound waves (see chapter 9). The idea of the presence of soft spots in the glass, regions of lower elastic modulus with respect to the surroundings, is common also to another approach that considers the BP as vibrations of nanometer sized regions[4]. A different approach, initially based on numerical calculations[5] and more recently formalized in a theoretical framework[6], is the fluctuating elasticity theory. The model is based on the idea that the microscopic elastic modulus of glasses has random spatial fluctuations and predicts the emergence of a boson peak and the Rayleigh scattering of the sound waves[7–10] (see chapter 10). This model assumes the validity of the harmonic approximation, an assumption common also to the random matrix theory[11]. Another family of models describes the boson peak within the jamming scenario[12], originally developed for a-thermal glasses, such as granular materials and colloids. Finally, we can mention a model where the BP is considered as a smeared first van Hove singularity, pushed down in frequency with respect to the crystal by hybridization effects[13].

Quite recently, numerical simulation studies (see chapter 11) have succeeded in the determination of the nature of the vibrational modes in the frequency range of the BP and below, employing systems composed of millions of atoms[14–19]. These studies suggest the presence of a non-negligible fraction of quasi localized excitations that coexist with the sound waves. However, these simulations are performed on model systems, based on quite simple interatomic potentials. Moreover, it has been shown that the density of quasi local modes decreases significantly in more stable glasses, which resemble more closely those obtained experimentally[19]. Consequently, the relevance of these findings for real glasses is still to be demonstrated.

This chapter presents a review of the main experimental observations on the vibrational dynamics of amorphous solids. We restrict our attention mainly to scattering experiments performed by means of neutron and X-rays. Section 2 is devoted to the results obtained by measurements of the density of vibrational states. These studies include the dependence of the boson peak on the thermodynamic parameters, such as temperature and pressure, and on the variation of macroscopic quantities, such as density and elastic moduli. The section is closed by a paragraph where the DOS of glasses is compared to the one of the corresponding crystals. Section 3 summarizes the main experimental observations on the wavevector dependence of the dynamic structure factor. These include the dispersion and broadening of the inelastic peaks, the Rayleigh scattering of the sound waves, evidences of transverse dynamics and the comparison with the dynamics of polycrystals.

## 2. Vibrational density of states

The vibrational density of states, $g(\omega)$, is a well-defined quantity both in crystals and in amorphous materials. It was recognized long ago[20], by means of neutron inelastic scattering, that the DOS is similar in glasses and in the corresponding crystalline form, as shown in Figure 1 for the case of vitreous silica (upper panel) and of $\alpha$-quartz (lower panel). At sufficiently low frequencies the DOS of crystals follows the Debye law, $g_D(\omega) = 3\omega^2/\omega_D^3$, where $\omega_D$ is the Debye frequency. However, the DOS of glasses differs from that of crystals in this low frequency range, below a few terahertz.

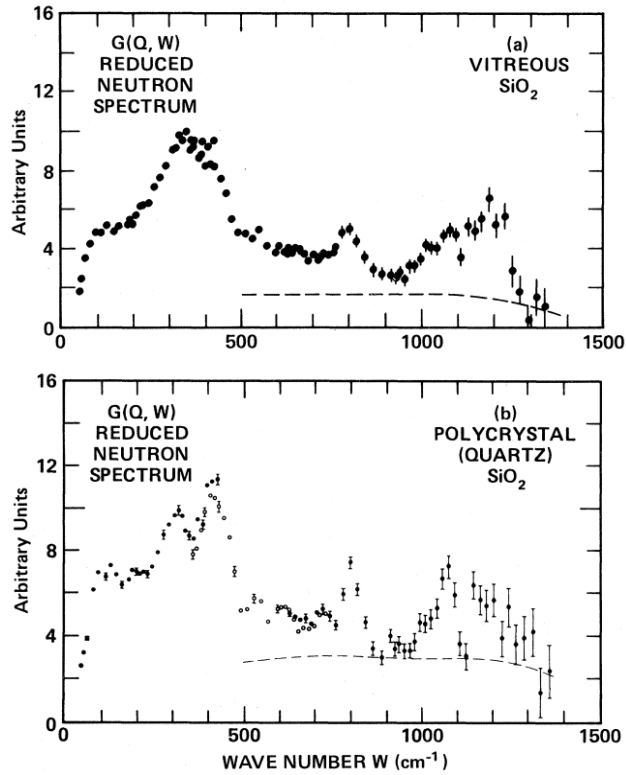

Fig. 1. DOS, $g(\omega)$, measured by neutron inelastic scattering in vitreous silica (upper panel) and in crystalline quartz (lower panel). Reprinted figure with permission from [F. L. Galeener, A. J. Leadbetter, M. W. Stringfellow, *Phys. Rev. B* **27**, 1052 (1983).][20] Copyright (1983) by the American Physical Society.

The corresponding vibrations are responsible for the well-known anomalous low temperature thermal properties of amorphous solids (see chapter 1 and 2 for details). The glass DOS does not extrapolate to the expected Debye behavior and the deviation is better appreciated by plotting the "reduced" DOS, $g(\omega)/\omega^2$. For the case of the canonical glass of silica this quantity extrapolates to a value which is more than twice the one expected from the Debye model (see Figure 2b) from Ref.[21]).

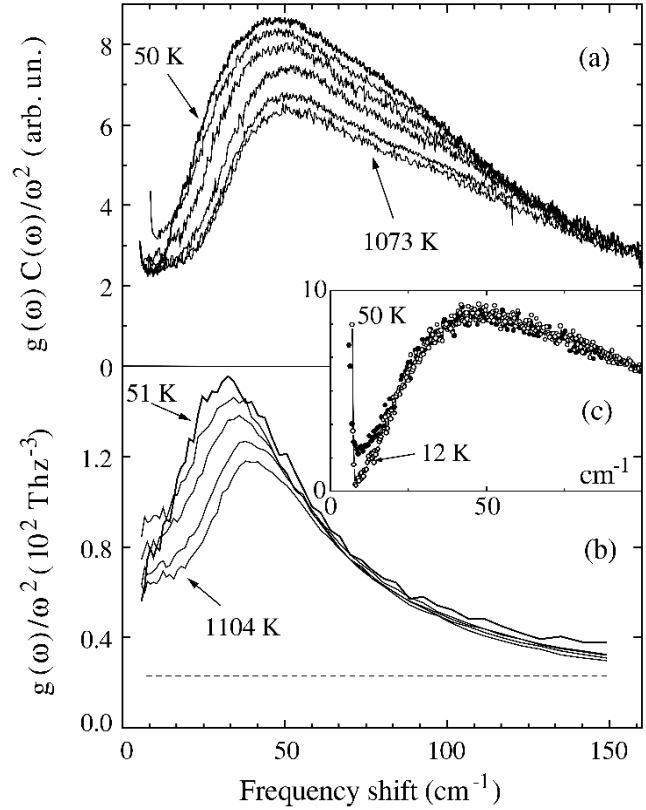

Fig. 2. Reduced density of vibrational states, $g(\omega)/\omega^2$, of vitreous silica as a function of temperature from Ref.[21] Panel a): Raman data (temperatures from top to bottom: 45, 323, 423, 523, 873, and 1073 K). Panel b): inelastic neutron scattering (from top to bottom: 51, 318, 523, 873, and 1104 K). The dashed line is the Debye level computed from the Brillouin sound velocities at 50 K. Panel c): Reduced Raman spectra at 12 and 50 K. All the temperatures are below the glass transition, which in $SiO_2$ is approximately 1450 K[22]. Reproduced from [A. Fontana *et al.*, *Europhys. Lett.* **47**, 56 (1999).][21]

Moreover, the reduced DOS presents a peak, the "boson peak". In most glasses the BP is located at frequencies close to one terahertz and corresponds to a peak in the specific heat over temperature cube, $C_p/T^3$, at a few degrees Kelvin. The example of the silica glass is shown in Figure 2, where Raman (panel a) and inelastic neutron scattering data (panel b) are compared. The name "boson peak" takes its origin from the fact that the peak in $g(\omega)/\omega^2$ is temperature independent at sufficiently low temperatures, as shown in Figure 2c) for silica. Consequently, in scattering experiments, both with electromagnetic waves and with neutrons, the intensity at the frequency of the BP follows the Bose-Einstein statistics, at least at sufficiently low temperatures. At higher temperatures the peak shifts, typically towards lower energies. The situation in silica is quite peculiar because this system becomes harder with increasing temperature and the BP correspondingly shifts to higher energies.

The BP in the reduced Raman scattering spectra (Figure 2a) and c)) is shifted to higher frequency with respect to its position in the neutron spectra because the light-vibrations coupling constant[23], $C(\omega)$, is approximately linear in $\omega$ around the BP[21,24]. It is worth mentioning that, already in their first seminal contribution[25], Buchenau and coworkers observed that the BP modes of vitreous silica were dominated by coupled rotations of corner sharing $SiO_4$ tetrahedra.

## 2.1. *Boson peak shape and intensity*

Malinovsky and coworkers showed that the BP shape is quite similar in different kind of solids, from network forming to metallic glasses, and suggested a universal shape for the excess DOS over the Debye level in the form of a logarithmic-normal function[26,27]. We have compared the reduced DOS of a set of ten amorphous solids, by plotting the different curves, all measured by neutron scattering, normalized to their maximum value and as a function of $\tilde{\omega} = \omega/\omega_{BP}$. The result is shown in Figure 3a) and suggests that the BP shape is not universal, although it is similar for a set of six out of the ten glasses considered. It should be mentioned, however, that this observation could be affected by the way the data are treated and by the fact that the spectra are collected from different neutron spectrometers. Keeping this in mind, we observe that $Se$[28], poly(isobutylene) (PIB)[29] and a metallic glass $(Ni_{33}Zr_{67})$[30] show a similar normalized reduced DOS in the chosen $\tilde{\omega}$ range, with an excess wing at high frequencies with respect to the other glasses.

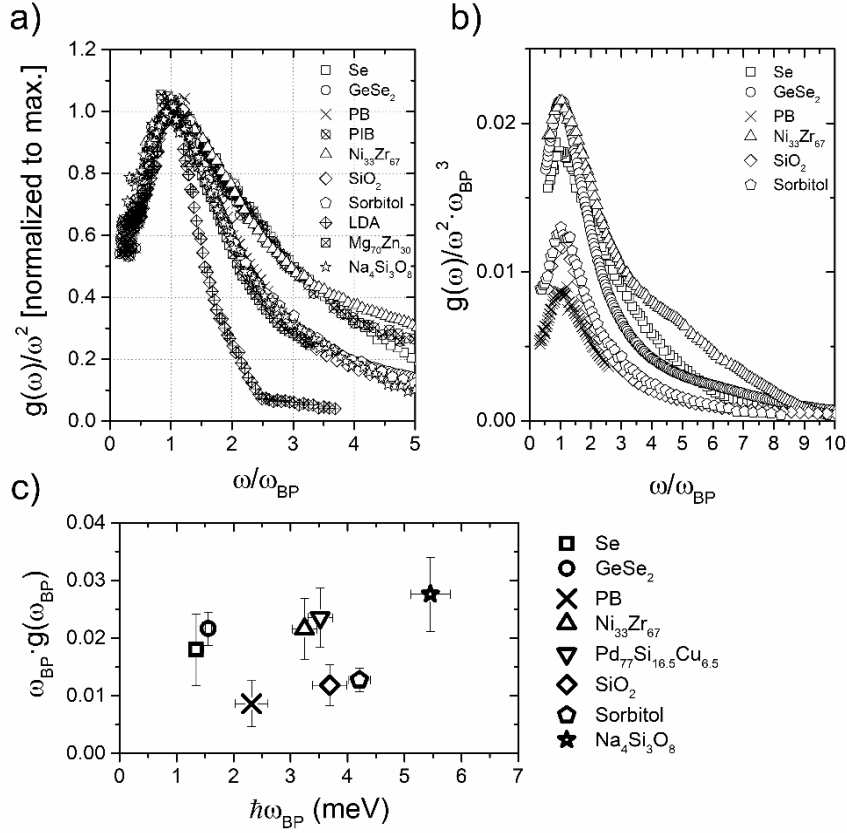

Fig. 3. Boson peak of a selection of glasses determined by neutron scattering: Se (squares)[28], GeSe$_2$ (circles)[31], polybutadiene (PB, crosses)[32], poly(isobutylene) (PIB, crossed circles)[29], Ni$_{33}$Zr$_{67}$ (up triangles)[30], Pd$_{77}$Si$_{16.5}$Cu$_{6.5}$ (down triangles)[33], SiO$_2$ (diamonds)[34], sorbitol (pentagons)[35], low density amorphous ice (LDA, crossed diamonds)[36], Mg$_{70}$Zn$_{30}$ (crossed squares)[26], 3(SiO$_2$)-2(Na$_2$O) (stars)[37]. Panel a): reduced DOS plotted as a function of $\omega/\omega_{BP}$ and normalized to the maximum. Panel b): reduced DOS scaled according to the mode conservation, eq. (1). Panel c): Maximum of the curves like those in panel b) for eight glasses as a function of the BP position.

On the contrary, low density amorphous ice[36] (LDA) has a reduced number of states at frequencies above the BP. The shape of the reduced DOS below $\omega_{BP}$ is instead quite similar in the ten systems considered. Since $g(\omega)$ counts the number of states, it was later recognized[38] that a rescaling of the ordinate axis to $\widetilde{\omega} = \omega/\omega_{BP}$ should be accompanied by a corresponding renormalization of the DOS to:

$$\widetilde{g}(\widetilde{\omega}) = g(\omega) \cdot \frac{d\omega}{d\widetilde{\omega}} = g(\omega) \cdot \omega_{BP} \qquad (1)$$

Despite the fact that the maximum of the reduced DOS has strong variations between the considered glasses, those differences almost disappear once the spectra are scaled by the BP position and the mode conservation of equation (1), which implies a multiplication by $\omega_{BP}^3$, is taken into account. The result is shown in Figure 3b) for a subset of the glasses considered in Figure 3a). The position of the maximum has been determined by fitting with a logarithmic-normal distribution a small portion around the peak. The BP intensity scaled in this way lies in the range 0.01 to 0.03 for all these materials. The precision of the scaling is affected by the uncertainty in the determination of the peak position and intensity. This is highlighted in Figure 3c) where we report the maximum of the rescaled reduced DOS as a function of the BP position. The maximum of $\widetilde{g}(\widetilde{\omega})/\widetilde{\omega}^2$ is simply the product of the BP frequency times the density of states at the BP, a quantity related to the integral of the DOS from zero to $\omega_{BP}$. This quantity appears to be almost independent of the BP position and with variations between different systems which are often within the uncertainty. Consequently, it seems that the number of vibrational states at the BP does not vary appreciably with the peak position and is unrelated to the Debye level, which is considerably different between the selected glasses.

Sokolov and coworkers observed that the ratio between the specific heat and the Debye expectation at the temperature of the BP maximum correlates well with the fragility of the glass, with a higher value for stronger glasses[39]. Because of the direct relationship between specific heat and DOS[40] this correlation implies a higher value of $g(\omega_{BP})/g_D(\omega_{BP})$ for a strong glass like silica (fragility index $\sim$20) and a lower one for a fragile glass like sorbitol (fragility index $\sim$100). The result of Figure 3c) suggest that the number of vibrational states is not strongly correlated with the glass fragility, which is instead probably correlated with the Debye level, at least for the systems considered in Ref. [39]

We can conclude that there is some evidence that the total number of vibrational modes at low frequencies, below the BP region, is not markedly different in different glasses. There is, on the contrary, a stronger variation in the BP position which can range from 1 to 10 meV[41]. However, at present there is no clear microscopic explanation of the value of $\omega_{BP}$, we only know that it is typically higher in systems with a higher value of the elastic modulus and there is some indication that the local structure plays a role[42,43].

## 2.2. *Boson peak dependence on external parameters*

Various studies have been performed with the aim to elucidate the dependence of the low frequency DOS on the variation of some external parameter, like pressure[29,44], temperature[38,43], sample density for permanently densified glasses[45–51], quenching rate[52], aging[53,54] or sample preparation by chemical vitrification[55]. All these studies indicate that the BP shifts to higher frequencies following an increase of the macroscopic elastic moduli. When the variation is small, below a few percent, the change of the DOS can be described by the elastic medium transformation, taking only into account the variation of the Debye frequency[52], as shown in Figure 4a) for the case of a silicate glass enriched in iron.

In the case of larger variations of $\omega_{BP}$, which are typically obtained by the application of pressure or by permanent densification, the modification of the DOS cannot be described by the macroscopic elastic medium transformation alone. Niss and coworkers, comparing neutron scattering and Brillouin measurements of poly(isobutylene) (PIB) as a function of applied pressure, observed a variation of $\omega_{BP}$ twice larger than that expected from the sound velocity values[29]. They showed that the shape of the reduced DOS was not affected by the application of pressure (see Figure 4b)), even if $\omega_{BP}$ at the highest applied pressure doubled its value at ambient conditions. Permanent densification of a silicate glass enriched in iron[48] indicated that the elastic medium transformation can account for the modification of the reduced DOS only if the local structure of the glass is not affected by the densification. In this material the first step of permanent densification induced some degree of modification of the local environment of the atoms, as suggested by Mossbauer spectroscopy, with the consequence that the variation of the BP shape from the normal density sample to the permanently densified ones could not be described by a simple elastic medium transformation. A subsequent work on permanently densified germanium dioxide[49] showed that the BP variation with density was higher than expected from the variation of the Debye energy. The reduced DOS of the permanently densified samples varied according to the mode conservation law of equation (1), as shown in Figure 4d). However, as in the previous example, the first step of densification induced some severe modification of the local structure which induced a strong change of the DOS. Equation (1) accounts well for the modification of the DOS in the case of the temperature evolution of the BP in vitreous silica (Figure 3c)), where it shifts from approximately 4 meV at low temperatures to 6 meV at the glass transition temperature.

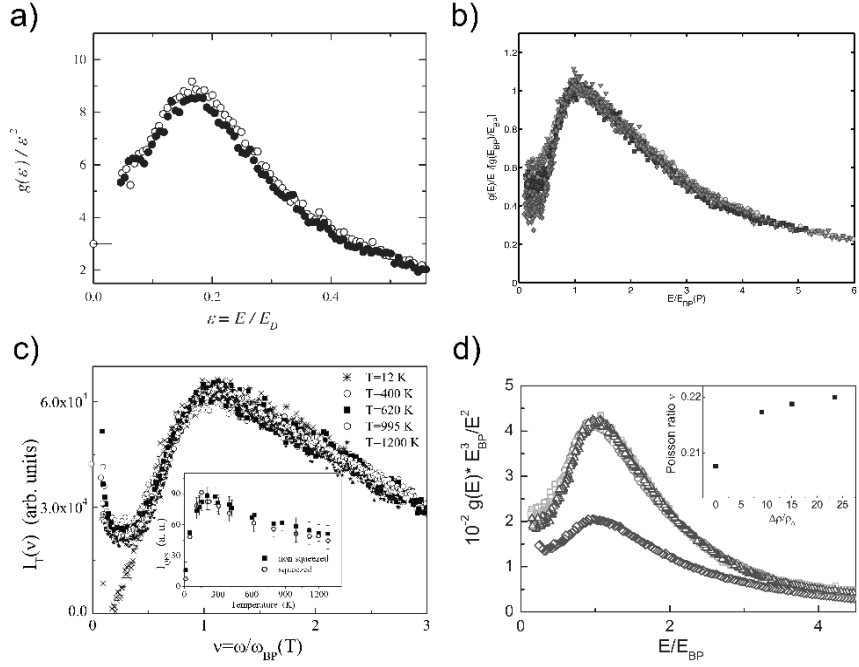

Fig. 4. Panel a): The difference in the BP of an hyperquenched (open symbols) and an annealed (solid symbols) silicate glass is well described by the elastic medium transformation, i.e. by the variation of the Debye frequency. Reprinted figure with permission from [A. Monaco *et al.*, *Phys. Rev. Lett.* **97**, 135501 (2006).][52] Copyright (2006) by the American Physical Society. Panel b): The shape of the reduced DOS of PIB is not affected by pressure, although the BP position changes by more than a factor of two. Data obtained by neutron scattering at: atmospheric pressure (dots), 0.4 (triangles), 0.8 (squares) and 1.4 GPa (diamonds). Reprinted figure with permission from [K. Niss *et al.*, *Phys. Rev. Lett.* **99**, 055502 (2007).][29] Copyright (2007) by the American Physical Society. Panel c): reduced Raman intensity of vitreous silica at the indicated temperatures rescaled following equation (1) and the hypothesis that the Raman coupling function is linear in frequency. The inset shows the Raman quasielastic intensity at 9.5 cm⁻¹. Reprinted figure with permission from [S. Caponi *et al.*, *Phys. Rev. B* **76**, 092201 (2007).][38] Copyright (2007) by the American Physical Society. Panel d): neutron scattering data of permanently densified GeO₂ glasses[49]. Normal glass (diamonds) and glasses permanently densified at the following pressures: 2 GPa (circles), 4 GPa (triangles) and 6 GPa (squares). The reduced DOS is rescaled following the mode conservation, eq. (1). The inset shows the Poisson ratio as a function of the relative density variation. Reproduced from [L. Orsingher *et al.*, *J. Chem. Phys.* **132**, 124508 (2010)][49], with the permission of AIP Publishing.

### 2.3. *Boson peak as a broadened van-Hove singularity*

In recent years, some inelastic scattering and specific heat measurements have revealed a significant connection between the boson peak of glasses and the first van Hove singularity present in the corresponding crystal. The most studied glass was silica[51] but other systems show a similar phenomenology, like the glassy mineral aegirine[42] and v-GeO$_2$[56]. The absence of periodicity characteristic of disordered systems such as glasses makes it difficult to describe their microscopic dynamics and the differences with respect to crystals pose several intriguing issues. In a crystal, in the long wavelength limit, the experiments agree with the Debye prediction while in glasses, as discussed in the previous paragraphs, this does not happen. The connection between the macroscopic elasticity and the microscopic vibrational dynamics of crystals is well understood. At the scale where the wavelength becomes comparable with the interatomic distances, the vibrational excitations gradually abandon their long-wavelength character, and this is marked by a bending-over of the acoustic branches, giving rise to the van-Hove singularities. This causes in crystals a deviation of the DOS from the low frequency Debye dependence. The boson peak in a glass and the lowest van-Hove singularity of the corresponding crystal are both expected to appear in the low frequency region where the sound waves experience a transition from the macroscopic continuum limit to the microscopic regime. However, it must be considered that usually the vitreous system and the corresponding crystal can have a notably different mass density. In fact, the lower density of glasses with respect to the corresponding crystals is a well-known consequence of the way glasses are prepared from the melt. The density dependence of the boson peak and its tendency towards the first van Hove singularity of the corresponding crystal has been demonstrated by nuclear resonant analysis of inelastic X-ray scattering[57] on the aegirine glass[42], by means of in-situ high pressure measurements.

This observation has been later confirmed[51] by comparing the DOS of permanently densified silica with the same density of crystalline $\alpha$-quartz and normal density silica with cristobalite. The experimental results are shown in Figure 5 where the DOS, $g(E)$, and the reduced DOS, $g(E)/E^2$, of $\alpha$-cristobalite, $\rho = 2.29$ g/cm$^3$, and $\alpha$-quartz, $\rho = 2.65$ g/cm$^3$, are compared with those of v-SiO$_2$, $\rho = 2.20$ g/cm$^3$, and permanently densified v-SiO$_2$, $\rho = 2.65$ g/cm$^3$, respectively. The $g(E)$ of glasses and of the corresponding crystals are very similar and differ mainly because glasses present a smoother and slightly broader DOS, with a stringent qualitative and quantitative correspondence. This similarity is even more evident if we look at the BP. In fact, the boson peak of v-SiO$_2$ is located nearly at the same energy of that of the van Hove singularity of $\alpha$-cristobalite. Concerning the comparison between the densified silica and $\alpha$-quartz, the energies are slightly different, but these small differences can be attributed to the distribution of the sizes of the local structures of the glass. The pseudo-Brillouin zone of glasses has a finite width that, once included in the crystal dispersion curves, can quantitatively account for the shift in the peak position as shown by the continuous curves in the figure and by the arrows. It is worth highlighting one detail, which can be appreciated in the bottom right panel, where the BP of densified silica is compared with the reduced DOS of $\alpha$-quartz. At low frequency the reduced DOS of the glass tends to a constant value, very close to that of the crystal and to the Debye level. Consequently, the densified silica is an example of a glass without appreciable excess of modes with respect to the Debye expectation. Other evidences are found also in boron oxide[58] and in glycerol[59,60].

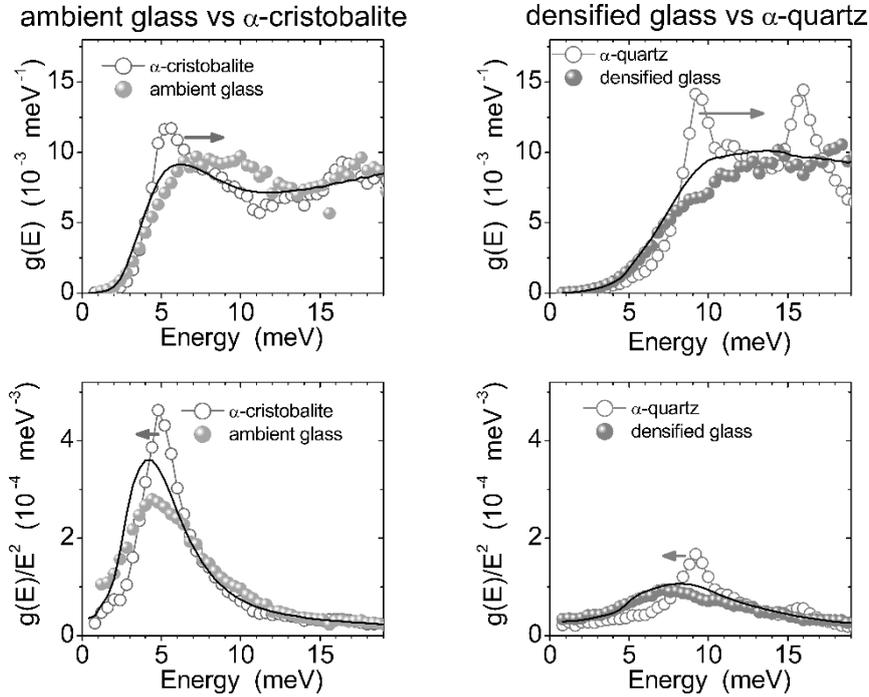

Fig. 5. Left: Comparison of the DOS (upper panel) and reduced DOS (lower panel) of ambient $SiO_2$ glass, full symbols, and of crystalline $\alpha$-crystobalite, open symbols. Right: DOS (upper panel) and reduced DOS (lower panel) of permanently densified $SiO_2$ glass, full symbols, and of crystalline $\alpha$-quartz, open symbols. The lines are obtained from the dispersion curves of the crystals including the effect of the smearing-out of the pseudo-Brillouin zone boundary given by the width of the first sharp diffraction peak of the glass. The arrows indicate the peak shift associated to this mechanism. Adapted with permission from [A. I. Chumakov *et al.*, *Phys. Rev. Lett.* **112**, 025502 (2014).][51] Copyrighted by the American Physical Society.

A similar observation, of a relation between the BP and the lowest energy van Hove singularity of the corresponding crystal, was done by Taraskin and coauthors in molecular dynamics (MD) simulations[13]. However, they interpreted the difference in energy between the BP of the glass and the first van Hove singularity of the crystal not in terms of a difference in mass density but as an effect of disorder that should shift, at the boundaries of the vibrational spectrum, the modes to lower energies.

## 3. Dispersion and scattering of vibrational modes

Even though the density of states provides many important information on the dynamics of disordered systems, as documented in the previous paragraphs, it cannot give conclusive results on the nature of the excitations. A richest wealth of information can be gained by studying the wave-vector resolved dynamical structure factor, $S(q, \omega)$. A variety of experimental techniques has been developed to probe this quantity in different regions of the $q - \omega$ space, but neutron or X-ray inelastic scattering are better suited to probe the terahertz frequency range of interest to our discussion.

Neutron scattering is the technique of choice for the study of the vibrational dynamics of crystals, where the kinematical restrictions imposed by the limited neutron energy can be avoided by probing the sample dynamics outside of the first Brillouin zone. Brillouin neutron scattering can be performed also in the first zone but only in samples with a sound velocity smaller than that of the incoming neutron. This limitation restricts the technique mainly to liquid samples, while in solid amorphous materials and in polycrystals the longitudinal vibrational modes in the first Brillouin zone is most often outside of the kinematical range, apart from specific glasses with low sound velocity[31,61,62]. X-ray inelastic scattering is a powerful method to study the dynamics of systems that do not possess translational periodicity, such as glasses, liquids, or polycrystalline materials, and as such it is complementary to neutron inelastic scattering.

We will provide in this section an overview of the main experimental results obtained on the investigation of the vibrational dynamics of glasses by probing the wave-vector resolved spectral distribution, $S(q, \omega)$, with these experimental methods.

### 3.1. *Dynamic structure factor*

Assuming for simplicity to deal with a monoatomic solid, the coherent cross section for both neutrons and X-ray scattering is proportional to the dynamical structure factor[63]:

$$S(q, \omega) = \frac{1}{2\pi N} \int_{-\infty}^{+\infty} dt\, e^{i\omega t} \sum_{l,l'} \langle e^{-iq \cdot \hat{r}_l(t)} e^{iq \cdot \hat{r}_{l'}(0)} \rangle, \quad (2)$$

where $\hat{r}_l(t)$ is the position operator of atom $l$ at time $t$ and $N$ is the number of atoms of the system. In a solid composed of more than one atomic species, the scattering cross section is proportional to a weighted average of the partial dynamical structure factors, where the weights are the appropriate scattering lengths for neutrons or the atomic form factors for the Thompson scattering of X-rays.

The dynamic structure factor is the sum of an elastic, $S^{(0)}(\boldsymbol{q}, \omega)$, and of an inelastic term, $S^{(in)}(\boldsymbol{q}, \omega)$. Assuming the atomic displacement of atom $l$, $\hat{u}_l(t)$, around the equilibrium position $\hat{x}_l$, to be sufficiently small we can write the elastic term as:

$$S^{(0)}(\boldsymbol{q}, \omega) = \frac{1}{N} \left| \sum_l e^{-i\boldsymbol{q} \cdot \hat{x}_l} e^{-\frac{W_l(q)}{2}} \right|^2 \delta(\omega), \qquad (3)$$

where $W_l(\boldsymbol{q}) = \langle (\boldsymbol{q} \cdot \hat{u}_l)^2 \rangle$ is the Debye-Waller factor for atom $l$. In a perfect crystal $S^{(0)}$ is different from zero only when the Laue condition is satisfied and $\boldsymbol{q}$ equals a reciprocal lattice vector. In a glass the structural disorder implies the presence of a non-negligible elastic component at all values of $q$. The inelastic term $S^{(in)}$ in the harmonic approximation is dominated by the one-phonon contribution:

$$S^{(1)}(\boldsymbol{q}, \omega) = \frac{\hbar}{Nm} \sum_\lambda \left| \sum_l e^{-i\boldsymbol{q} \cdot \hat{x}_l} e^{-\frac{W_l(q)}{2}} \boldsymbol{q} \cdot \boldsymbol{e}_l(\lambda) \right|^2 \text{x} \qquad (4)$$

$$\frac{1}{2\omega_\lambda} \{ [n(\omega_\lambda, T) + 1]\delta(\omega - \omega_\lambda) + n(\omega_\lambda, T)\delta(\omega + \omega_\lambda) \}.$$

In this expression $m$ is the mass of the atom, $\boldsymbol{e}_l(\lambda)$ is the $\lambda$-th eigenvector at atom $l$ of eigenfrequency $\omega_\lambda$ and $n(\omega, T)$ is the Bose population factor for vibrations of frequency $\omega$ at temperature $T$. Positive frequencies in equation (4) correspond to the Stokes side where the peak intensity of the ground state ($T = 0$) is non-zero.

For an ideal crystal the $S^{(1)}$ is peaked at frequencies corresponding to the dispersion curves of phonons with a wave-vector $\boldsymbol{k}$ satisfying the relation $\boldsymbol{q} = \boldsymbol{k} + \boldsymbol{G}$, where $\boldsymbol{G}$ is a reciprocal lattice vector. Consequently, the dynamic structure factor of a crystal is composed of a set of sharp peaks, which are broadened by defects and anharmonicities. In cases where the measurement is confined within the first Brillouin zone, the one-phonon term includes contributions coming only from the longitudinal component of the eigenmodes. The nature of the eigenmodes, which cannot be associated to a well-defined $\boldsymbol{k}$ vector, of a disordered solid implies $S^{(1)}$ to be composed of broad peaks even in the harmonic approximation. The broadening comes from the sum over the eigenvalues $\lambda$ in equation (4).

The experimental inelastic X-ray scattering (IXS) spectra are typically analyzed in terms of one or more damped harmonic oscillation (DHO) functions that describe the inelastic part of the spectrum. The single DHO model for the dynamic structure factor is[64]:

$$S(q, \omega) = S(q)\{f_q \delta(\omega) + \left(1 - f_q\right) \frac{\hbar \omega}{k_B T}[n(\omega, T) + 1] \text{ x} \qquad (5)$$

$$\text{x} \frac{1}{\pi} \frac{\Omega^2(q) \Gamma(q)}{[\omega^2 - \Omega^2(q)]^2 + \omega^2 \Gamma^2(q)}\},$$

where $S(q)$ is the static structure factor. The parameters of the model are: the non-ergodicity parameter, $f_q$, defined as the ratio of elastic to the total intensity, the position of the peak in the longitudinal current, $\Omega(q)$, and the parameter $\Gamma(q)$ that estimates the full width at half maximum of the inelastic peak.

This simple model is useful to analyse the experimental spectra, which are often measured at a fixed $q$ as a function of frequency. However, the model itself finds a proper justification only in the hydrodynamic regime of simple liquids. In the case of solids or highly viscous systems it is conveniently used to determine the positions and widths of the inelastic features, but it has clear limitations, in particular at high $q$ values. This will be shown in the next paragraph.

## 3.2. *Single damped harmonic oscillator analysis*

### 3.2.1. *Dispersion of the inelastic peak position*

An example where the single DHO model works well is reported in Figure 6 for the case of glassy glycerol. The spectrum presents a marked elastic line whose width is fixed by the instrumental response function, since possible relaxation processes contributing to the width of the central line are undetectable with the available energy resolution of approximately 1.5 meV.

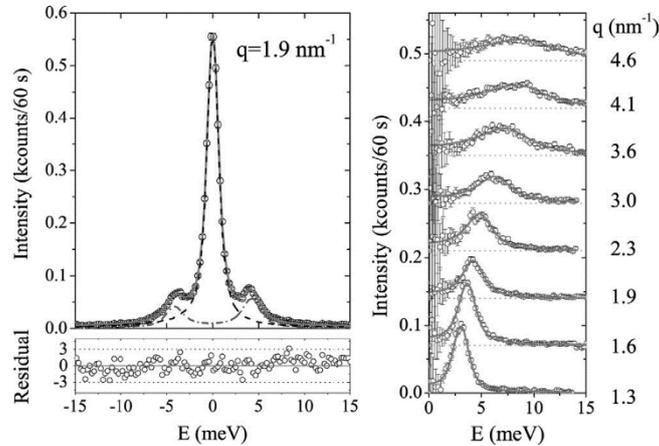

Fig. 6. Left column: Inelastic X-ray scattering spectrum of glassy glycerol at the indicated exchanged wavevector together with the best fitting function (continuous line), modelled as an elastic line (dashed) plus a damped harmonic oscillator describing the inelastic component (dot-dashed). See text for details on the data analysis. The bottom panel shows the residues between the model and the data. Right column: inelastic intensity on the Stokes side at selected $q$s, together with the best fitting functions. The elastic line is subtracted after the fit is performed on the entire spectrum. Reproduced from [G. Monaco and V. M. Giordano, *Proc. Natl. Acad. Sci.* **106**, 3659 (2009).][65]

The inelastic features are composed of a Stokes and an anti-Stokes peak and present some dispersion as a function of $q$, as highlighted in the right panel of the figure. Here the Stokes side is plotted after subtraction of the elastic line, whose intensity is determined by a fit of the entire spectrum. The validity of the DHO model as a proper description of the glass spectra has been debated in the literature[66–68]. However, it is now commonly accepted as a useful fitting model, better suited to describe the spectra in the low $q$ region, far from the position of the first sharp diffraction peak (FSDP) in the static structure factor. As the wavevector is increased, the DHO model has the tendency to overestimate the inelastic intensity at the centre of the spectrum[69]. The dispersion of the Brillouin peak has been documented in a variety of systems, from oxides to organic and metallic glasses[31,33,35,37,41,61,62,65,66,68–84]. A few examples are reported in Figure 7.

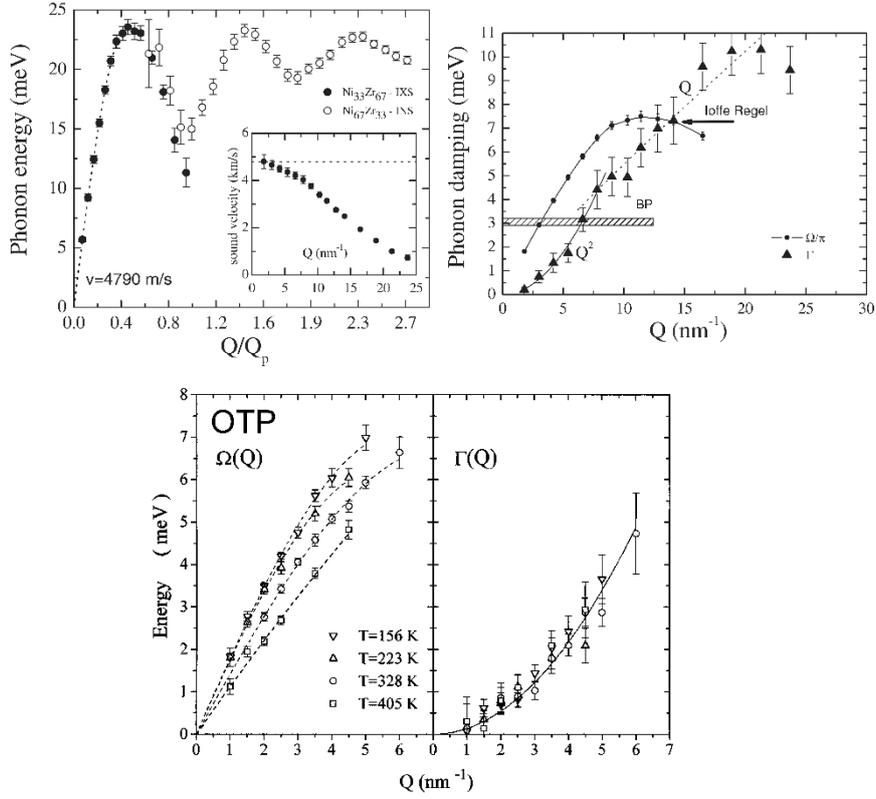

Fig. 7. Top: Dispersion (left) and damping (right) of the inelastic peak of a metallic glass. The left panel compares data for $Ni_{33}Zr_{67}$ measured by IXS[85] with those of $Ni_{67}Zr_{33}$ investigated by neutron inelastic scattering (INS)[86], scaling the exchanged wavevector by the first sharp diffraction peak position, $Q_P$. ($Q_P^{Ni_{33}Zr_{67}} = 25.7$ nm$^{-1}$ and $Q_P^{Ni_{67}Zr_{33}} = 29$ nm$^{-1}$). Inset: Wavevector dependence of the apparent sound velocity of $Ni_{33}Zr_{67}$ probed by IXS. The phonon damping of $Ni_{33}Zr_{67}$ is shown in the right panel (triangles) together with the dispersion curve plotted as $\Omega/\pi$ (dots). The intersection between the two data sets defines the Ioffe-Regel limit, here observed in the second pseudo-Brillouin zone, much above the boson peak frequency (horizontal bar). Reprinted figures with permission from [T. Scopigno *et al.*, *Phys. Rev. Lett.* **96**, 135501 (2006).][85] Copyright (2006) by the American Physical Society. Bottom: Dispersion (left) and damping (right) of OTP as a function of temperature across the glass transition ($T_g = 243$ K). Reprinted figure with permission from [G. Monaco *et al.*, *Phys. Rev. Lett.* **80**, 2161 (1998).][71] Copyright (1998) by the American Physical Society.

The dispersion of the peak position as a function of $q$ in a metallic glass[85] is shown in the upper left panel of the figure, together with the apparent sound velocity shown in the inset. At low wave-vectors the dispersion is linear and follows the sound velocity of the longitudinal acoustic mode. At higher $q$s the dispersion resembles that of a crystalline material, with a sinusoidal behaviour suggesting the presence of a pseudo-Brillouin zone even in a disordered structure. The zone boundary, $Q_P$ in the figure, has a value which is approximately one half of the FSDP position. The presence of a certain degree of periodicity of the dispersion curve has been sometimes interpreted in terms of Umklapp scattering[87]. Metallic glasses are characterized by a FSDP located at relatively high wavevectors and typically sharper than that of other glasses, a consequence of the dominantly radial metallic bonding. The clear signature of a pseudo-Brillouin zone and a periodicity which persists even in the fourth zone are most probably consequences of the somehow larger residual order in their disordered structure with respect to network forming, organic or polymeric glasses.

Clear evidences of a dispersion curve are observed in almost all the studied glasses, at least within the first pseudo-Brillouin zone. An example is the orto-terphenyl glass (OTP)[71] shown in the bottom panel of Figure 7. However, the width of the inelastic peak, $\Gamma$, of OTP becomes quickly comparable to the peak position as the border of the pseudo-Brillouin zone is approached (bottom part of the figure, right panel). When this happens, the inelastic feature cannot be considered as a damped vibrational mode but more properly as the spectral distribution of vibrational modes at the selected $q$. A common observation is the reduction of the sound velocity as the temperature is increased, reflecting the solid to liquid transition. On the contrary, the peak broadening is typically found to be temperature independent within experimental uncertainty[88], suggesting the predominantly harmonic nature of the vibrational modes, although evidence of an-harmonic damping has been reported in silicate glasses[76].

### 3.2.2. *Peak broadening and the Ioffe-Regel limit*

In many glasses the broadening of the inelastic peak is found to grow quadratically with the exchanged wavevector. This is the case in both systems of Figure 7. The metallic glass shows $\Gamma \sim q^2$ in the low $q$ region, followed by an approximately linear dependence at higher $q$s. This linear regime is a peculiarity of metallic glasses, reminiscent of the behavior of liquid metals[89]. The OTP glass shows a quadratic increase of $\Gamma$ with $q$ in the entire range shown in the bottom panel of Figure 7. The precise nature of this quadratic dependence is still not clear, despite its observation goes back more than twenty years. For the specific case of vitreous silica, we have a quite robust interpretation, which stems from the comparison with the corresponding crystalline structure[90]. We will discuss this point in more detail in the following paragraphs. A quadratic increase of the peak broadening with frequency can also be a consequence of an-harmonicities, as we observed in a sodium silicate glass[76].

It is commonly accepted that the inelastic peak can be considered as a damped longitudinal acoustic phonon if the broadening is much smaller than the peak position. A qualitative criterion to assess whether the peak can be treated as a single eigenmode is the Ioffe-Regel (IR) limit[91], defined by the condition:

$$l \sim \lambda. \tag{6}$$

The IR limit is reached at the frequency where the amplitude mean free path, $l$, equals a wavelength, $\lambda$. The condition can be rewritten in terms of the full width at half maximum, $\Gamma$, and of the position, $\Omega$, of the inelastic peak as:

$$\Gamma \sim \Omega/\pi. \tag{7}$$

It was initially believed, in analogy with the theory of electron conduction, that the IR limit marked the transition from propagating to localized excitations. That this is not the case became evident from MD simulations, showing that the vibrational modes above the IR limit have a high participation ratio, characteristic of extended vibrational modes[92]. The localization threshold is found at much higher frequency, towards the end of the spectrum (for instance, in silica at approximately 30 THz, or 1000 cm$^{-1}$, see Figure 1).

The IR limit in many glasses, including oxides, polymers and organic ones, has been found at a frequency comparable to that of the BP[41]. This coincidence between $\omega_{IR}$ and $\omega_{BP}$ has been debated in the literature[93–95]. In particular, it is not observed in metallic glasses [33,85]. An example is the NiZr of Figure 7, where the IR frequency is almost an order of magnitude higher than the BP frequency, as shown in the upper right panel of the figure. This observation further indicates how the dispersion curves of metallic glasses are well defined, with sharp inelastic features persisting at $q$ values significantly higher than in other materials.

### 3.2.3.  *Rayleigh-like scattering and negative dispersion*

Improvement in the energy resolution, luminosity and available flux of the IXS beamlines has allowed the detection of Rayleigh-like scattering of elastic waves in glasses[35,41,65,69,76,78,96,97]. In analogy with the scattering of electromagnetic radiation by small particles, localized impurities in a solid can cause a sound attenuation which increases as the fourth power of the frequency[98,99]. Recent theories for the vibrations in disordered solids suggest that Rayleigh scattering should be present also in glasses[10]. An increase of the sound damping as the fourth power of the wavevector has been indeed measured by means of IXS in a few systems, including vitreous silica[69], sodium silicate glasses[76], alkali borates[41], glycerol[65] and an organic glass[35]. In all these systems the Rayleigh scattering regime is measured at low frequencies, below the IR limit and is accompanied by a negative dispersion of the apparent sound velocity.

The data for silica at a temperature slightly above the glass transition are reported in Figure 8. The sound velocity, top panel, has an initial decrease from the macroscopic value of the longitudinal sound wave followed by an increase at frequencies exceeding the boson peak position. The negative dispersion typically accompanies the Rayleigh scattering, while the positive dispersion is not always observed. The BP frequency at this temperature is highlighted by the vertical dashed line and marks the upper frequency for the Rayleigh scattering, as shown in the middle panel. Also shown is the coincidence between $\omega_{IR}$ and $\omega_{BP}$. At higher frequencies, the width of the inelastic peak, estimated by a DHO model, follows a quadratic frequency dependence, like those seen in Figure 7. The bottom panel of the figure shows the reduced DOS as a function of temperature, highlighting the shift of the BP position as the temperature is increased, a peculiarity of vitreous silica whose elastic modulus increases with temperature (DOS from neutron scattering as in Figure 2, bottom panel, but in a wider T range).

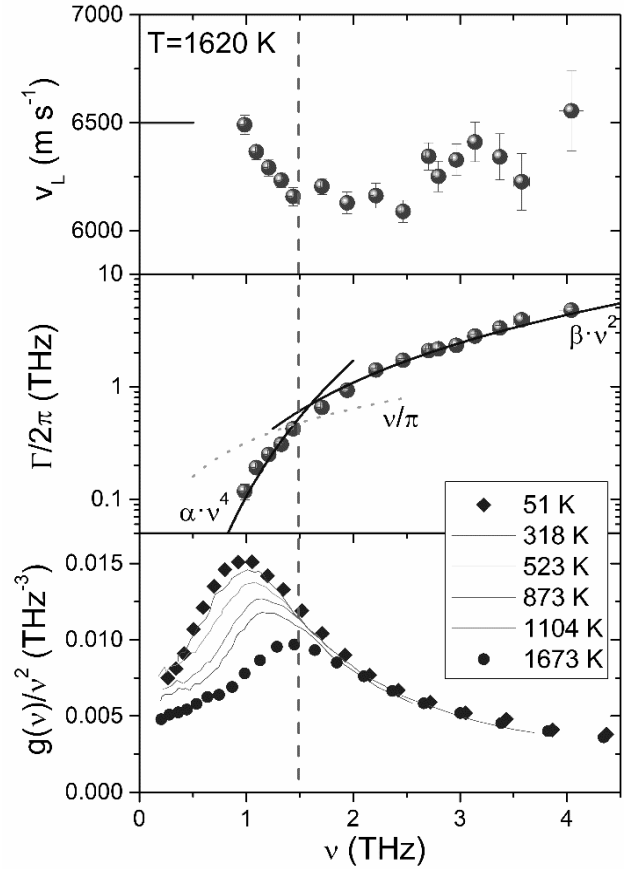

Fig. 8. Inelastic X-ray (the two upper panels) and neutron scattering (lower panel) data of v-SiO₂. The IXS data are collected at 1620 K, above the glass transition temperature. The upper panel shows the apparent sound velocity as determined from the ratio of the inelastic peak position and the exchanged wavevector. The peak broadening increases as the fourth power of the frequency (Rayleigh scattering) below the boson peak position and quadratically above. The two power-law dependences are indicated by the continuous lines. The vertical dashed line indicates the BP position determined by INS at a similar temperature (bottom panel)[21,100]. The Ioffe-Regel frequency perfectly matches the BP frequency in this system, as indicated by the dashed line in the middle panel showing the quantity $\nu/\pi$. Reprinted figure with permission from [G. Baldi et al., *Phys. Rev. Lett.* **104**, 195501 (2010).][69] Copyright (2010) by the American Physical Society.

The negative dispersion evidenced by the reduction of the sound velocity in the low frequency part of the figure (upper panel) indicates that the Debye approximation is no longer valid at these frequencies. A deviation from the linear dispersion of long wavelengths sound waves has a direct effect on the density of states, giving rise to a deviation from the Debye law. However, a quantitative evaluation of the contribution of the observed negative dispersion to the DOS requires an estimate of the dispersion of the transverse acoustic modes, which are not directly accessible experimentally, at least in this low frequency regime. Assuming that the same negative dispersion affects also the transverse branch one can estimate the reduced DOS, which is in reasonable agreement with the boson peak position and intensity both in glycerol and in sorbitol glasses[35,65]. A more precise estimate is obtained by including the broadening of the longitudinal and transverse branches. However, the small departure from a linear dispersion is not sufficient to account for the BP intensity in the case of silica and silicate glasses[69,76]. In silica the agreement between the dispersion curve and the DOS can be recovered by considering the bending of the acoustic-like branch towards the border of the pseudo-Brillouin zone[51], in analogy with the van-Hove singularity of the corresponding crystal, as discussed in section 2.3.

This bending is not observed in the analysis reported in Figure 8, where the inelastic features are described in terms of a single excitation model. Evidences of a more complex phenomenology, with at least two separate inelastic peaks, is found in the high $q$ range and will be discussed in section 3.3.

### 3.2.4.  *Sound attenuation in glasses*

Rayleigh scattering persists up to the boson peak frequency, above which the peak width follows a different law, growing with the square of the frequency in most glasses (see Figures 7 and 8). The microscopic origin of the Rayleigh scattering in glasses is debated. One interpretation is found within the soft potential model[101]. The model (see more in chapter 9) assumes the existence of quasi-localized vibrational modes, which should coexist with the sound waves in the low frequency range, below the BP. Interaction between the sound waves and the quasi localized modes is predicted to give rise to the strong scattering, with the damping growing as the fourth power of frequency[101]. This view has received some support very recently, because MD simulations on model systems composed of a huge number of particles allow one to probe with accuracy the low frequency part of the DOS, revealing that indeed a considerable fraction of the vibrational modes are quasi localized in this frequency range[16,18]. However, it has also been shown that the density of quasi localized modes depends on the way the glass is prepared. Specifically, more stable glasses tend to have a much lower density of quasi-localized, non-phononic, excitations with respect to glasses obtained by a fast quench[19,102]. Since the experiments are typically performed on stable glasses, it is difficult to judge at present if this mechanism is indeed the one responsible for the Rayleigh scattering.

A different view considers Rayleigh scattering in glasses to arise from local elastic modulus fluctuations[5,7] (see more in chapter 10). These elastic heterogeneities would strongly scatter the sound waves, giving rise to the $\Gamma \sim q^4$ law. The strong scattering regime is observed when the wavevector is sufficiently small that the condition: $qa \ll 1$ is satisfied, where $a$ is the mean size of the elastic modulus heterogeneities. In this view the BP would correspond to that frequency where the wavelength satisfies $\lambda \sim 2\pi a$ and the sound wave starts to feel the microscopic heterogeneity of the glass.

The IXS technique can probe the Rayleigh scattering only in a quite restricted frequency window, mainly because of limitations in the available energy resolution. A more detailed investigation of the sound attenuation in glasses requires the use of multiple experimental probes, including ultrasonic techniques, Brillouin light (BLS)[103,104] and UV (BUVS) scattering[105,106], pico-second optical methods (POT)[107] and the tunneling junction (TJ) technique[108]. A summary of some of the data available for vitreous silica is reported in Figure 9. The frequency range between 100 GHz and 1 THz is the most difficult to probe. This region can be investigated, at present, only with sophisticated ultrasonic techniques based on the generation of acoustic pulses with optical methods[109–111].

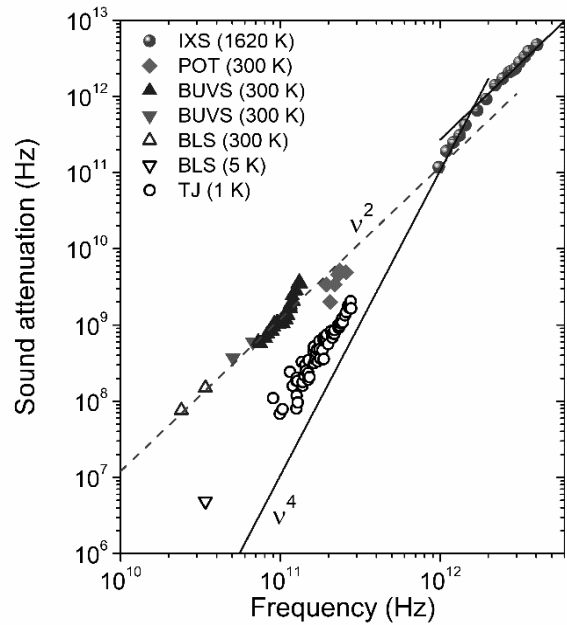

Fig 9. Sound attenuation in vitreous silica as a function frequency at the temperatures indicated in the legend. Collection of data obtained with various techniques: inelastic X-ray scattering (IXS)[69] above $T_g$; picosecond optical technique (POT)[107], Brillouin ultraviolet scattering (BUVS) with synchrotron radiation[106] (upper full triangles) and with a laser source[105] (lower full triangles). Brillouin light scattering (BLS)[103,104] and tunneling junction method (TJ)[108]. Reprinted figure with permission from [G. Baldi *et al.*, *Phys. Rev. Lett.* **104**, 195501 (2010).][69] Copyright (2010) by the American Physical Society.

Another possible approach, which has been recently demonstrated, is the transient grating spectroscopy with deep UV radiation produced by a free electron laser[112]. However, both approaches can probe the sound attenuation of longitudinal waves only in thin films of a thickness of a few tens of nanometers, contrary to the bulk samples that can be studied with Brillouin scattering of light or X-rays.

The continuous line in Figure 9 is an extrapolation of the Rayleigh scattering observed by IXS towards lower frequencies, suggesting that the sound damping probed by the lower frequency methods is dominated by other attenuation mechanisms. Specifically, the room temperature data follow a quadratic frequency dependence, typical of an-harmonic and relaxational contributions. More recent data[109–111] confirm this trend up to approximately 300 GHz. However, also the data at lower temperatures remain higher than the expected Rayleigh term. Room temperature IXS data, not shown in the figure, give some evidence that this discrepancy could be associated with the variation of the BP position with temperature[113]. The Rayleigh scattering seems to be slightly shifted to lower frequencies at room temperature, following the shift of the boson peak, in qualitative agreement with the TJ data at 5 K. Nevertheless, the room temperature IXS data are quite noisy and it is difficult to draw a definite conclusion from the experiments.

Recently, a series of numerical works has observed the possible existence of a logarithmic correction to the Rayleigh law, suggesting that the discrepancy between the TJ and the IXS data in silica is indeed related to this correction[114]. Those numerical works indicate the origin of the logarithmic correction in a long-range correlation of the elastic modulus fluctuations[115]. Other MD works associate this anomalous Rayleigh scattering to the presence of the previously mentioned quasi localized vibrational modes[102].

### 3.3. *Evidences of two excitations*

The existence of acoustic-like excitations propagating up to terahertz frequencies in all the amorphous solids studied so far has stimulated the search for evidences of transverse acoustic modes[31,33,61,62,82,116–120]. Numerical simulations[82,95,121–123] suggest the boson peak to have a predominantly transverse character, stimulating research in this direction. In a purely acoustic picture, the weight of the transverse modes in the DOS far exceeds that of the longitudinal ones because of their lower sound velocity, as in the Debye model. An X-ray or neutron scattering experiment is, however, blind to the transverse vibrational modes of an isotropic solid, being sensitive only to vibrations with a component parallel to the exchanged wavevector $q$, as discussed in section 3.1. Indeed, only a single inelastic peak is measured, at least for wavevectors below one-half of the pseudo-Brillouin zone boundary. At shorter wavelengths - higher $q$s - a second inelastic feature appears in the spectra and becomes progressively more and more evident as $q$ increases, as reported by Ruzicka et al.[116] in a first X-ray scattering study on silica.

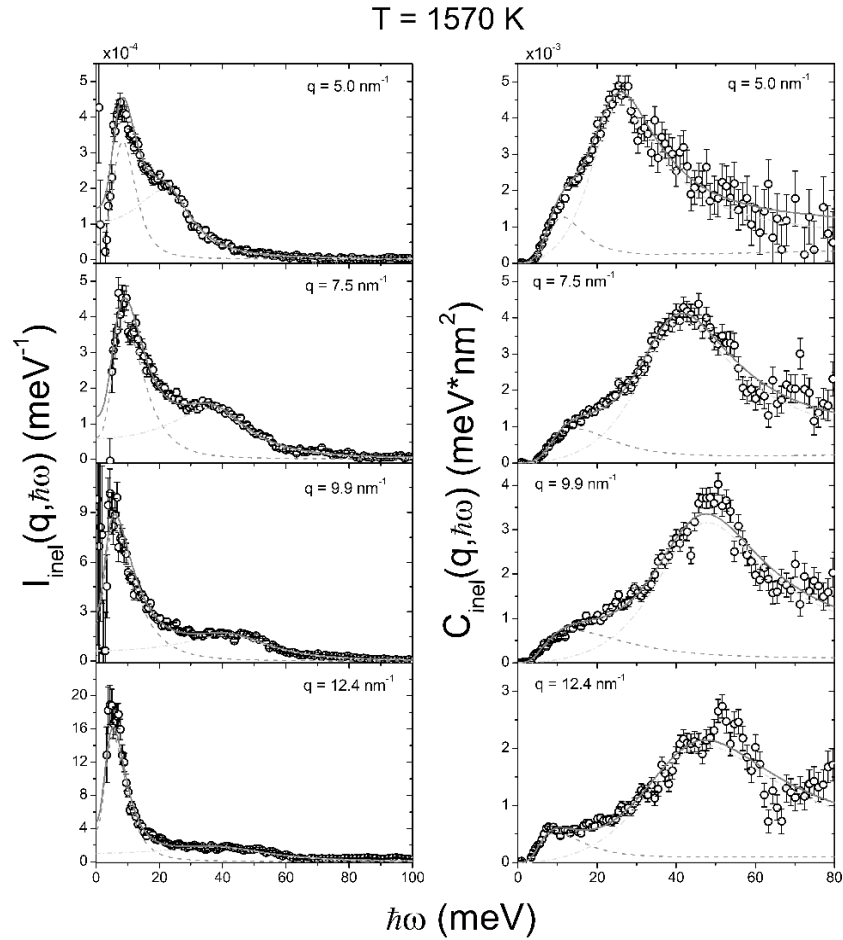

Fig. 10. Inelastic part of the IXS spectra measured at T=1570 K on v-SiO$_2$ at selected exchanged wave vectors; for the details of data treatment see Ref.[119]. In the left panel the spectra are plotted as $I_{inel}(q, \omega)$. The continuous line is the best-fitting function obtained by means of two DHO functions (dashed and dashed-dotted lines) convoluted with the instrumental resolution function. The right panel shows the same spectra plotted as longitudinal inelastic currents; same symbols as in the left panel. Reprinted figure with permission from [G. Baldi *et al.*, *Phys. Rev. B* **77**, 214309 (2008).][119] Copyright (2008) by the American Physical Society.

The inelastic scattering intensity together with the corresponding current correlation functions of v-$SiO_2$ are shown in Figure 10 at selected $q$ values, as measured in a later experiment[119]. The inelastic spectra show two evident broad features. The lower-frequency inelastic component is better evidenced in the plots of the inelastic intensity, $I_{inel}(q, \omega)$, left column of the figure, while the higher frequency one is enhanced in the current correlation function, $C_{inel}(q, \omega)$, where $C_{inel} = \omega^2/q^2 \cdot I_{inel}$, by the $\omega^2$ factor. Evidences of two inelastic features appearing within the first pseudo-Brillouin zone have been reported for systems of different structure and composition both with IXS and INS. Examples include other oxide glasses[82], metallic glasses[33,80], chalcogenides[31,61], organic materials[62,117,120] and even water[118]. The case of the metallic glass $Pd_{77}Si_{16.5}Cu_{6.5}$ is included in Figure 11.

The two inelastic features observed in v-$SiO_2$ present a dispersion with $q$, as shown in the upper panel of Figure 11, left side, together with points determined in previous investigations and including neutron data for the lower frequency peak. It is worth noting that data previously shown in the upper panels of Figure 8 span the wavevector range from approximately 1 to 4.5 nm$^{-1}$. The macroscopic longitudinal and transverse sound velocities measured by Brillouin light scattering[124] are used to calculate the linear dispersion curves shown in the figure. Two inelastic features are present in the spectra when $q$ exceeds some 4 nm$^{-1}$. The first few points of the lower frequency excitation are in good agreement with the linear dispersion of the macroscopic transverse sound wave, suggesting that the second excitation which appears in the spectrum is indeed the transverse acoustic mode. More precisely, this mode corresponds to the projection of the transverse acoustic mode in the longitudinal current, which is not negligible because of the high value of the wavevector. The phenomenon is sometimes called "mixing" and was initially revealed in numerical simulation studies where it appeared clear that the wave polarization was progressively lost as the wavevector increased towards the pseudo-Brillouin zone boundary[122,125].

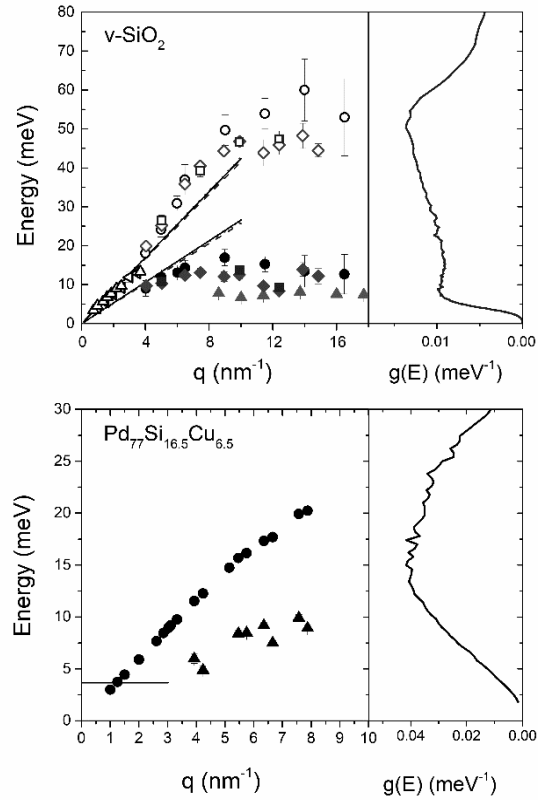

Fig. 11. Top. Left side: dispersion relation of v-SiO₂ from the IXS and INS data. Open (longitudinal branch) and filled (transverse branch) diamonds: T=1570 K, open and filled squares: T=920 K, from ref.[119]; open and filled circles: Ref. [116], T =1270 K; up triangles: from Ref. [126], T=1375 K; down triangles: Ref. [66], T=1050 K. The position of the peak in the INS spectra is reported as filled triangles[119]. The lines correspond to the macroscopic sound velocity measured by Brillouin light scattering[124] for the longitudinal and transverse sound waves: $v_L$=6320 m/s and $v_T$=3910 m/s at T =920 K (dashed), $v_L$=6440 m/s and $v_T$=4040 m/s at T=1270 K (continuous), $v_L$=6500 m/s and $v_T$=3980 m/s at T=1570 K (dotted). Reprinted figure with permission from [G. Baldi *et al.*, *Phys. Rev. B* **77**, 214309 (2008).][119] Copyright (2008) by the American Physical Society. Right side: $g(\omega)$ measured by INS, revised data from Ref.[34]. Bottom. Left side: dispersion relation of the Pd₇₇Si₁₆.₅Cu₆.₅ metallic glass from IXS measurements. Longitudinal branch (full circles) and transverse branch (full triangles). The straight line indicates the position of the boson peak. Right side: $g(\omega)$ measured by means of INS. Adapted from [P. Bruna *et al.*, *J. Chem. Phys.* **135**, 101101 (2011)][33], with the permission of AIP Publishing.

In fact, the comparison of the transverse and longitudinal current correlation functions, calculated from MD simulations, indicated that, on increasing $q$, the transverse peak appeared also in the longitudinal current and, vice-versa, the longitudinal peak appeared in the transverse one. It should also be kept in mind that the two inelastic features are broad peaks since they appear at a wavevector well above the Ioffe-Regel limit. Consequently, the two inelastic features cannot be considered as two excitations but more properly as a distribution of many eigenstates. It is also of interest to consider the energies of the peaks of the two inelastic features. The acoustic modes of quartz and cristobalite are restricted to below some 20 meV. Above this energy the vibrations of the crystalline polymorphs of silica are optic modes. This observation suggests that the lower frequency inelastic peak of vitreous silica has a predominantly acoustic character, while the higher frequency one is mostly equivalent to the optic modes of the corresponding crystal. The dispersion of the higher frequency peak displays, in fact, a strong deviation from the linear dispersion of the longitudinal branch just above the point where the second excitation appears in the spectrum. This kind of "positive dispersion" is possibly related to a partial mixing of acoustic and optical branches, whose presence in the glass is confirmed by Raman, neutron scattering and infrared absorption measurements[20,127]. Both inelastic features lose any appreciable dispersion above 6-7 $nm^{-1}$. This $q$ value is close to the pseudo-Brillouin zone boundary estimated as one-half the position of the first sharp diffraction peak in the static structure factor.

The dispersion curves of the metallic glass $Pd_{77}Si_{16.5}Cu_{6.5}$ are shown in the lower panel of Figure 11. A second, lower frequency excitation, appears also in this system when $q$ exceeds approximately one-half of the pseudo-Brillouin zone boundary. The DOS, $g(\omega)$, obtained by inelastic neutron scattering at room temperature on v-$SiO_2$ is reported in the right side of the top panel of Figure 11, to compare it with the dispersion curves. The comparison between the DOS and the dispersion curves shows that the dynamic structure factor evolves towards the density of vibrational states as $q$ is increased, as expected from the incoherent approximation, whose validity is assumed in the calculation of $g(\omega)$ from the neutron spectra.

In the case of vitreous silica, the figure indicates that the lower frequency inelastic feature evolves towards the lower peak of $g(\omega)$, located at approximately 10 meV. When seen as reduced DOS, $g(\omega)/\omega^2$, this first peak is the boson peak and is located at 4 meV. The higher frequency inelastic feature of the silica glass evolves towards a second peak, appearing in the DOS at about 50 meV. The convergence of the lower frequency peak towards the transverse acoustic dispersion at low $q$s confirms that at least a part of the BP intensity has a transverse nature. This discussion does not exclude the presence of additional modes in the low frequency part of the spectrum, such as the quasi localized excitations observed in recent numerical simulations[16–19]. However, the presence of those modes is not revealed in scattering experiments as the main contribution to the Brillouin scattering, even at terahertz frequencies, comes from the acoustic-like excitations discussed above.

### 3.4. *Dynamics of glasses and corresponding polycrystals*

Scattering experiments carried out within the first pseudo-Brillouin zone indicate that the dynamic structure factor evolves as a function of $q$ and displays oscillations reminiscent of crystals dispersion curves. However, in most systems, with the notable exception of metallic glasses, the Ioffe-Regel limit is reached at a relatively low frequency, comparable to the position of the boson peak. At least in these glasses we cannot thus talk of vibrational branches, where well-defined excitations follow a dispersion curve. On the contrary, the experiments suggest the presence of a distribution of vibrational modes whose amplitude has a non-trivial $q$-dependence, gradually evolving towards that expected from the density of states. Consequently, scattering experiments do not tell us much about the nature of the glasses vibrational modes at frequencies exceeding the Ioffe-Regel limit. At lower frequencies, on the contrary, the Brillouin spectra clearly indicate the presence of a longitudinal acoustic mode with a sound velocity compatible with the one determined from macroscopic measurements.

A line of research has tried to compare the dynamics of the glass with that of the corresponding crystal, to obtain some additional information on the nature of the terahertz vibrations of glasses, in the same spirit of the approaches discussed in section 2.3.

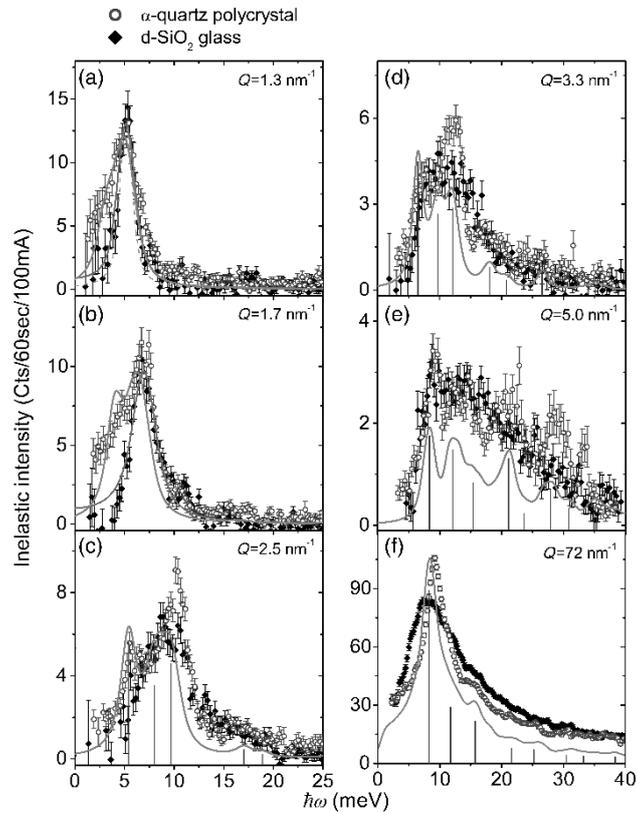

Fig 12. Comparison of the inelastic spectra, Stokes side, of an $\alpha$-quartz polycrystal (open circles) and of permanently densified $SiO_2$ glass (full diamonds) with matching mass densities. The spectra are measured by means of IXS and shown at selected wave-vectors. The continuous line is the IXS spectrum of a polycrystal of $\alpha$-quartz determined with an ab-initio calculation. The vertical lines indicate the average positions of the different branches of the crystal. The three lines of lower frequency derive from the acoustic branches while the higher frequency ones correspond to the optic branches of the single crystal. It is worth noting that the spectra are reported in absolute units, without any rescaling. Panel a) and b) include also the fit of the glass spectrum with a DHO model. The instrumental resolution function is reported as a dashed line in panel a). Reprinted figure with permission from [G. Baldi *et al.*, *Phys. Rev. Lett.* **110**, 185503 (2013).][90] Copyright (2013) by the American Physical Society.

The idea is to compare a glass with the crystalline polymorph with similar macroscopic properties, such as sound velocities and mass density. Unfortunately, it is not always possible to perform this comparison either because of difficulties in preparing the crystal or because of the presence of more than one crystalline structure preventing the identification of the one that corresponds more closely to the glass, in terms of macroscopic elastic properties. Consequently, the number of studies is still quite limited and includes investigations performed on ethanol[74], amorphous water[72], vitreous silica[128], $GeO_2$[129], $BeF_2$[129] and permanently densified silica[90].

We will here focus our attention to the case of the permanently densified silica with a mass density close to that of the crystal of $\alpha$-quartz. The study has been performed by means of IXS and has investigated in detail the low $q$ region, including the range extending from the typical macroscopic response up to the border of the pseudo-Brillouin zone[90]. The Stokes side of the inelastic spectra of the glass and of $\alpha$-quartz polycrystal are compared in Figure 12 at selected $q$ values together with the spectrum of the polycrystal determined by an ab-initio calculation. A striking result is the observation of two regimes. A low $q$ regime where the inelastic spectra of the polycrystal and of the glass are different and a high $q$ region where the spectra of the two samples are very similar. The transition between the two takes place at a crossover wavevector $Q_c \sim 2$ nm$^{-1}$. In the low $q$ region the spectrum of the polycrystal is broader than that of the glass, while above the crossover wavevector the inelastic spectra of the two become almost indistinguishable. The polycrystal spectrum presents sharper features with respect to that of the glass at all the probed $q$s, with peaks that correspond to the single crystal branches averaged over all the possible directions in reciprocal space. In the limit of high $q$s the incoherent approximation applies, and the spectrum resembles the density of states (panel f of Figure 12).

The distinction between the two regimes, above and below $Q_c$, is better highlighted by fitting the spectrum with a single excitation model, a simple damped harmonic oscillator. The DHO model allows one to give a rough estimate of the width of the spectrum, even when it is composed by a distribution of normal modes and not by a single excitation. The result of this analysis is shown in Figure 13. Below $Q_c$ the glass spectrum is well described by the single DHO model, as shown in Figure 12, panels a) and b). The mode can be identified as a longitudinal acoustic vibration because its dispersion follows the longitudinal sound velocity.

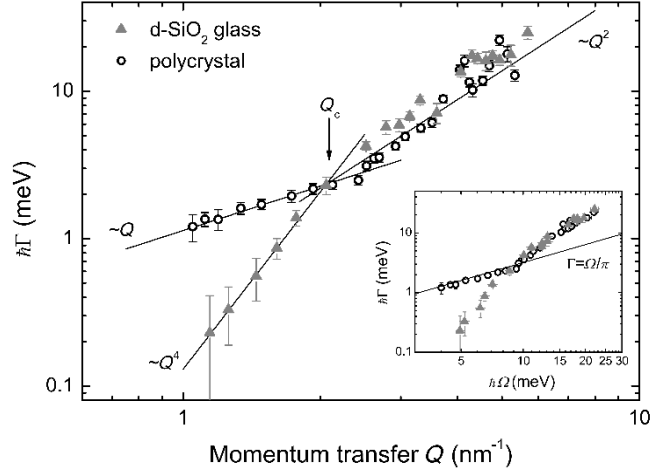

Fig. 13. Full width at half maximum of the DHO model used to fit the spectra of a permanently densified silica glass (full triangles) and of $\alpha$-quartz polycrystal (open circles). The lines indicate different power law behaviors. $Q_c$ is the crossover wavevector above which the spectra of the glass and the polycrystal become very similar. Inset: same data as a function of the frequency of the inelastic peak (the maximum of the DHO, which corresponds to the maximum of the current correlation function). The line indicates the Ioffe-Regel condition. Reprinted figure with permission from [G. Baldi *et al.*, *Phys. Rev. Lett.* **110**, 185503 (2013).][90] Copyright (2013) by the American Physical Society.

The sound attenuation follows the Rayleigh scattering law and increases rapidly with $q$ until the spectrum becomes comparable to that of the polycrystal at $Q_c$. In this low $q$ region the spectrum of the polycrystal is broader than that of the glass and its width follows a linear dependence on $q$. The ab-initio calculation well reproduces the experimental spectrum and shows that this linear increase of the linewidth is a consequence of the contribution of the quasi-transverse branches, whose polarization is not purely transverse because of the anisotropy of quartz. At wavevectors above $Q_c$ the peak broadening for both the glass and the polycrystal follows a quadratic dependence on $q$. This growth for the polycrystal is due to the quadratic increase with $q$ of the optic mode intensity, whose effect is a corresponding spectral broadening.

It is worth noting that the Ioffe-Regel condition coincides within experimental uncertainty with the linear broadening of the polycrystal spectrum, as shown in the inset of Figure 13. This apparent coincidence can be understood as follows. The mode dispersion in this frequency range is approximately linear and the ab-initio calculation suggests the contribution of all three acoustic branches. If, for simplicity, we assume to have only one quasi-transverse and one quasi-longitudinal branch with a similar weight, the peak broadening is expected to be:

$$\Delta\omega \sim (\mathrm{v}_L - \mathrm{v}_T)q \sim \left(1 - \frac{\mathrm{v}_T}{\mathrm{v}_L}\right)\omega \sim \frac{\omega}{3}. \tag{8}$$

Here $\mathrm{v}_L$ and $\mathrm{v}_T$ are the macroscopic longitudinal and transverse sound velocities, with $\mathrm{v}_T \sim 2/3 \cdot \mathrm{v}_L$ for silica. Condition (8) is very similar to the Ioffe-Regel limit, equation (7). Moreover, the transition between the macroscopic regime for the glass, where a single acoustic mode has a damping which follows the Rayleigh law, and the microscopic range, where the spectra of the polycrystal are very similar to those of the glass, takes place at the boson peak energy, at about 9 meV (Figure 12 f)). Since the boson peak has approximately the same position of the first van-Hove singularity of the crystal, the proximity of the Ioffe-Regel and the BP frequency is possibly a coincidence, related to the shape of the dispersion of the quasi-transverse branch.

The comparison of the dynamics in glasses and in the corresponding polycrystals can help to shed some light on the nature of the glass vibrational modes in this low frequency regime, at least for the case of densified silica. One can imagine the glass elasticity to be heterogeneous at the nanometer or sub-nanometer scale, the typical length being set by $1/Q_c$. For wavelengths exceeding $2\pi/Q_c$ the glass behaves as an isotropic continuum and the fluctuations of the local elastic modulus give rise to the Rayleigh scattering. At small wavelengths, $\lambda < 2\pi/Q_c$, the elastic wave feels the local atomic structure and the dynamics is similar to the one of the corresponding polycrystal. However, because of the high anisotropy of quartz, it is difficult to make predictions on the nature of the vibrational modes of the polycrystal, because the crystal symmetry is lost by the average in reciprocal space and many eigenmodes contribute to the spectrum. Nevertheless, the similarity between the polycrystal and the glass spectrum suggests that the modes around the boson peak frequency are predominantly extended and not localized. This is one of the main conclusions of the studies here summarized, together with the estimate of the average size of the local elastic modulus fluctuations, $\sim 1/Q_c \sim 0.5$ nm.

Extensions of this approach to other glasses are not trivial, both because of limitations in the wavevector range which is experimentally accessible and because of the difficulty of finding the proper corresponding polycrystal of a given glass. At the same time there are indications of counterexamples where the dynamics of the glass and of the corresponding crystal do not match. This discrepancy has been reported in a two-dimensional model system[130] and in numerical simulations of a Lennard-Jones glass[14].

## 4. Conclusions

We have reviewed the main experimental observations of the vibrational dynamics of amorphous solids performed by means of scattering methods. The measurement of the density of vibrational states indicates, in all the studied glasses, the presence of a boson peak, whose position spans a range of approximately a decade in frequency, centered around one terahertz. The shape of this feature is similar in very different glasses, particularly at frequencies below the peak. At present, the connection between the frequency of the BP and other parameters of the glass, such as its fragility or its elasticity, remains elusive. When the glass is subjected to variations of external parameters, such as pressure, density, temperature or annealing time, the BP has a shift and a corresponding variation of intensity, that normally follows a simple counting law, related to the normalization of the density of states. Applying the same counting law to a series of different glasses and scaling all the density of states by the BP position, we observe that the peak intensity has fluctuations to within a factor two or three but with considerable uncertainties.

A few studies have highlighted the similarity of the BP with the first van-Hove singularity of the corresponding crystal with comparable mass density and elastic properties. This equivalence is quite convincing in strong network forming glasses. Another interesting result is that the density of states of a few systems, densified silica[51], boron oxide[58] and glycerol[59], converges towards the Debye limit at low frequencies. These findings pose new questions on the role of the structural disorder as origin of the boson peak. Nevertheless, their generality should be verified by new scattering measurements on a wider selection of glasses.

Evidences of pseudo-acoustic modes propagating at terahertz frequencies in glasses have been provided by inelastic X-ray scattering experiments. The technique detects, in many glasses, an acoustic mode that follows, at low frequencies, the macroscopic longitudinal sound velocity. In those systems where the measurements can be performed at frequencies lower than the BP position, this acoustic mode has an attenuation that follows the Rayleigh scattering law. In the same frequency range the mode shows a negative dispersion, which can explain quantitatively the BP intensity in the cases of glycerol and sorbitol glasses[35,65]. In silicate glasses the observed negative dispersion alone is not sufficient to justify the deviation from the Debye law observed at the BP[76,113]. A quantitative comparison of the DOS of the glass with that of the corresponding crystal of similar density suggests that the BP is equivalent to the first van-Hove singularity[51,90]. However, the acoustic mode becomes ill-defined above the Ioffe-Regel limit, so that the nature of the glass vibrational modes around the BP frequency remains unclear.

When the exchanged wavevector exceeds the value where the inelastic spectrum is peaked at the BP frequency, the spectrum starts to display a complex phenomenology. We have reviewed the evidences of transverse dynamics mixed to the longitudinal branch and the appearance of two inelastic features. In a few cases, the wavevector evolution of the inelastic spectrum can be compared with that of the corresponding polycrystal. The dynamics is different below the BP frequency, where the glass shows a single damped pseudo-acoustic mode as expected for an isotropic solid. Above the BP the spectra of the glass and of the corresponding polycrystal become very similar, indicating the contribution of a multitude of eigenvectors. In the polycrystal these modes are associated with the various branches of the single crystal averaged over the directions in reciprocal space.

Recent numerical simulations of model systems composed of millions of atoms have highlighted the coexistence of phonon like modes with quasi localized excitations in glasses in the proximity of the BP frequency[16,17,19,102]. Although the applicability of these results to real systems is still under scrutiny, they motivate the realization of new experiments searching for evidences of these excitations. Another interesting issue regards the origin of the Rayleigh scattering and whether it is related to those quasi local modes or whether it originates from the elastic modulus fluctuations present in glasses at the nanometer scale. Recent theories for the elasticity in glasses suggest these fluctuations to extend to the long range, with the consequence that the Rayleigh law should be modified by a logarithmic correction[114,115,131,132]. Finally, the nice agreement between the spectra of the glass and of the corresponding polycrystal observed in a few systems seems questionable for the Lennard Jones glass, leaving the connection between the BP and the first van Hove singularity of the corresponding crystal an open question.